# Ferroelectricity Going 2D

Chuanshou Wang[†], Lu You[#], David Cobden[‡], Junling Wang[†§*]

[†]Department of Physics, Southern University of Science and Technology (SUSTech), Shenzhen, China

[§]Guangdong Provincial Key Laboratory of Functional Oxide Materials and Devices, Southern University of Science and Technology, Shenzhen 518055, Guangdong, China

[#]Jiangsu Key Laboratory of Thin Films, School of Physical Science and Technology, Soochow University, Suzhou 215006, China

[‡]Department of Physics, University of Washington, Seattle, USA

*jwang@sustech.edu.cn

## Abstract

The discoveries of magnetism and ferroelectricity in 2D van der Waals (vdW) materials have brought important functionalities to the 2D materials family, and may trigger a revolution in next generation nanoelectronics and spintronics. In this perspective article, we briefly review the recent progress in the field of 2D ferroelectrics, focusing on the mechanisms that drive spontaneous polarizations in 2D systems, unique properties brought about by the reduced lattice dimensionality, and promising applications of 2D ferroelectrics. At the end, we provide an outlook for challenges that need to be addressed and our view on possible future research directions.

## Main

A century has passed since the discovery of ferroelectricity in Rochelle salt [1]. We have learned a great deal about the fascinating physics of these materials, and developed a host of practical devices based on their switchable polarization and functionalities arising from it, such as ferroelectric random-access memory, actuators, transducers, infrared detectors and electro-optic modulators [2]. Previous studies have mostly focused on three-dimensional (3D) ferroelectric materials. For integration with the semiconductor industry, thin films are often grown on lattice matching substrates [3] and free-standing films can be prepared by using sacrificial layers [4]. However, the chemical and structural discontinuities at the surface/interface not only hinder investigations of the fundamental effects of reduced dimensionality, but also pose challenges for applications in electronics.

Recent developments in two-dimensional (2D) vdW ferroelectrics introduce a new paradigm to the field. Their layered structure allows for stable monolayer and few-layer samples [5], providing an excellent platform for studying the effects of reduced lattice dimensionality on long-range ferroelectric order. In just the last five years many 2D ferroelectric systems with different mechanisms behind the spontaneous polarization have been discovered, unique properties differing from those of 3D ferroelectrics have been uncovered, and functional devices making use of the 2D nature have been demonstrated. It is thus timely to survey the experimental findings and what has been learned, and to identify questions and challenges that have arisen. Here our intention is not to provide a detailed technical review, but rather to formulate a framework within which the existing results can be analyzed. Building upon the recent progress, we also provide our perspective on future directions in this field.

**Review of 2D ferroelectrics**

The study of 2D ferroelectrics can be traced back to investigations in the 1960s and 70s of group-IV semiconducting monochalcogenides (MX, where M=Ge/Sn; X=S/Se/Te) [6] and transition metal thio/selenophosphates (TPS, $[M^{I2+}]_2[P_2X_6]^{4-}$ and $M^{I1+}M^{III3+}[P_2X_6]^{4-}$, where $M^I$= Mn/Zn, $M^{II}$= Ag/Cu, $M^{III}$=Cr/V/In/Sc, X=S/Se) [7]. Early works were mostly conducted on bulk samples. Later with the development of thin film growth by molecular beam epitaxy (MBE) and more recently the introduction of mechanical exfoliation, ultra-thin samples of 2D ferroelectric materials became available. Robust out-of-plane polarization in 4 nm (~5 layer) $CuInP_2S_6$ [8] and in-plane polarization in monolayer SnTe [9] were reported in 2016. Since then, ferroelectricity has been demonstrated in many 2D systems including $In_2Se_3$ [10, 11], bis(benzylammonium) lead tetrachloride (BA$_2$PbCl$_4$) [12], 1T-WTe$_2$ [13-15], d1T-MoTe$_2$ [16], SnSe, SnS [17], bilayer graphene/h-BN heterostructure [18], twisted bilayer h-BN [19, 20], GeTe [21], rhombohedrally stacked bilayer WSe$_2$, MoSe$_2$, WS$_2$ and MoS$_2$ [22, 23], CuCrP$_2$S$_6$ [24] and NiI$_2$ [25,26] etc. Importantly, these 2D ferroelectric materials span the spectrum of electronic properties from insulators with large band gaps to semimetals with no band gap, and most have Curie temperatures (T$_C$s) close to or above room temperature which is crucial for practical applications.

## Origin of spontaneous polarization in 2D ferroelectric systems

**Ionic displacement induced polarization**. In perovskite ferroelectric oxides such as BaTiO$_3$, the hybridization between O *2p* and Ti *3d* states pushes B site ions away from the geometric centers of the oxygen octahedra, making the crystal lattice non-centrosymmetric and giving rise to the spontaneous polarization [27]. In monolayer SnTe, a representative group-IV monochalcogenides, the polarization comes from the relative intralayer displacement between Sn and Te ions, distorting the lattice from cubic to rhombohedral [9, 28]. Similarly, in CuInP$_2$S$_6$, the Cu ions shift away from the centers of the S octahedra [8, 29] (Figure 1a). Note that the ferroelectric phase transition of SnTe is of the pure displacive type, where the ionic displacement vanishes above T$_C$ [6], whereas CuInP$_2$S$_6$ undergoes an order-disorder phase transition at T$_C$ [29]. Here we emphasize only the origin of the spontaneous polarization. Such ionic displacement induced polarization also gives rise to the ferroelectricity in monolayer In$_2$Se$_3$ [10, 11] and 1dT-MoTe$_2$ [16, 30].

**Polarization from polar molecular groups**. In certain molecular crystals and polymers such as polyvinylidene fluoride (PVDF), the molecular units possess permanent dipole moments whose spontaneous ordering produces the polarization [31]. At the ferroelectric-paraelectric phase transition, the long-range order of the molecular dipoles is destroyed but the dipoles themselves survive. This is different from the situation in displacive-type ferroelectrics in which the dipoles vanish above the T$_C$. Recent rapid development of organic-inorganic hybrid materials has led to a myriad of hybrid ferroelectrics, many of which possess 1D and 2D structures [32]. However, ionic displacement and polar molecular group induced ferroelectricity are not mutually exclusive. Some of the molecular ferroelectrics share the specificities of both mechanisms, and so are some inorganic ferroelectrics [33]. For example, in the 2D hybrid ferroelectric $BA_2PbCl_4$, the polarization comes from both the ordering of organic molecular dipoles and the relative displacement between the negatively charged $PbCl_6$ octahedron and positively charged organic molecular groups. Together they give rise to a net in-plane polarization that can be switched by an external electric field [12] as shown in Figure 1b. Another example is the $CuInP_2S_6$, whose polarization comes from ionic displacement but which undergoes an order-disorder phase transition [29].

**Charge redistribution induced polarization**. In some 2D materials, theory suggests that interlayer charge redistribution through hybridization between the occupied states of one layer and the unoccupied states of the neighboring layer could induce out-of-plane electric dipoles [18-20, 23, 34]. A small relative shift between adjacent layers can convert the structure into its mirror image and reverse the charge redistribution [14]. This kind of ferroelectric switching was first observed in natural few-layer $WTe_2$ [13] and more recently has been demonstrated in artificially stacked bilayer h-BN [19, 20], as well as $WSe_2$, $MoSe_2$, $WS_2$ and $MoS_2$ [22, 23]. A similar phenomenon has also been reported in small-angle twisted bilayer graphene/h-BN moiré heterostructure [18]. When the artificially stacked bilayers are not perfectly aligned, as is usually the case, a moiré pattern of alternating polarization domains occurs [34] as shown in Figure 1c. It is also worth noting that in these systems the interlayer translation may be the leading order parameter rather than the polarization, making them improper ferroelectrics, though whether this is the case remains to be established.

**Spin-driven polarization**. Ferroelectricity can also be induced by certain kinds of long-range magnetic order, as in type-II multiferroics [35]. A notable example is perovskite manganite orthorhombic $TbMnO_3$. It is antiferromagnetic with a non-colinear spin structure, generating a net electric polarization via the inverse Dzyaloshinskii–Moriya interaction [36]. In 2D magnetic systems, it was predicted theoretically that spin-driven ferroelectricity may occur in $Hf_2VC_2F_2$ as illustrated in Figure 1d [37]. Recently spin-helix order was reported to induce finite electric polarization in monolayer $NiI_2$, which can be controlled by the spin chirality [25,26]. Additionally, magnetic order induced ferroelectricity was also demonstrated in vdW antiferromagnetic $CuCrP_2S_6$ due to spin-dependent *p–d* hybridization [24]. The intimate coupling between electric and magnetic orders in 2D multiferroics presents opportunities to observe new physical phenomena as well as potential for multifunctional devices.

## Unique properties due to the reduced lattice dimensionality

**Polar stability in ultra-thin films.** In thin films of 3D ferroelectric materials, chemical and structural disruption at a surface or interface often result in a so-called "dead layer", hampering polar stability in the ultrathin limit [38, 39]. In contrast, 2D materials can be easily exfoliated into monolayers without any dangling bonds. In fact, it has been reported that the $T_C$ of monolayer SnTe increases to above 270 K, much higher than the bulk value of 100 K, and the polarization also increases [9] as shown in Figure 2a. This anomalous behavior was attributed to two factors: first, quantum confinement enlarges the band gap and reduces the screening by free carriers; and second, the in-plane lattice distortion increases with decreasing thickness, making the polarization more stable [9].

**Electromechanical coupling in 2D ferroelectric systems.** An important application of ferroelectric materials is in actuators and transducers, which make use of the electromechanical coupling provided by the piezoelectric and converse piezoelectric effects. The weak interlayer coupling in vdW 2D ferroelectrics naturally leads to totally different electromechanical behavior. In typical 3D ferroelectric materials, when an electric field is applied along the polarization direction, the positive ions will move along the field direction, and the negative ions move oppositely, and this tends to expand the lattice and gives a positive longitudinal piezoelectric coefficient $d_{33}$ [40]. In a simple rigid ion model, this originates from anharmonicity due to the spontaneous ionic displacement (polarization). However, in a multilayer 2D ferroelectric with out-of-plane polarization, such as $CuInP_2S_6$, the dipoles are within separate layers. As an applied electric field strengthens the dipoles within each layer, the resulting enhanced dipole-dipole interaction causes the vdW gap to decrease by more than the expansion of the layers, leading to a net decrease in the overall lattice constant and negative $d_{33}$ (Figure 2b) [41, 42]. This model also explains the negative piezoelectricity found in PVDF [43] and predicted in BiTeI [44]. Furthermore, it is expected that a negative $d_{33}$ will occur ubiquitously in 2D (and 1D) ferroelectric materials (multilayers) with spontaneous polarization along the direction of the lattice disruption.

**Ferroelectric/Polar metals**. Ferroelectricity and metallicity are usually incompatible due to the screening of long-range Coulomb interaction between dipoles by high-density itinerant electrons. However, Anderson and Blount in 1965 pointed out the possibility of a "ferroelectric/polar metal" in which a polar axis may emerge during a symmetry-breaking structural transition in a metal [45]. This phenomenon was recently experimentally demonstrated in 3D $LiOsO_3$ and $NdNiO_3$ [46, 47]. Nevertheless, despite the polar distortion, screening by free carriers eliminates any net electrical polarization and prevents switching between the symmetry-broken states by an external field, so these polar metals are not strictly ferroelectrics. However, this is not the case for perpendicular polarization in 2D materials, where the carriers are only free to screen in the plane. Indeed, electric-field controlled out-of-plane polarization switching has been demonstrated in few-layer $WTe_2$ while it is in a metallic state, using graphene/h-BN/$WTe_2$/h-BN/graphene devices as shown in Figure 2c [13, 15]. It is believed that the confinement of itinerate electrons by the vdW gap plays a significant role [14]. The same principle should allow the existence of other ferroelectric metals and even ferroelectric 2D superconductors [48].

## Applications of 2D ferroelectric materials

**2D ferroelectric/multiferroic tunnel junctions.** In a ferroelectric tunnel junction (FTJ), the switchable polarization modulates the tunnelling energy barrier, resulting in tunneling electroresistance ratios (TERs) of ~10 to $10^6$ in traditional 3D ferroelectric (e.g. $BaTiO_3$, $Pb(Zr,Ti)O_3$, $HfO_2$ and $Hf_{0.5}Zr_{0.5}O_2$) heterostructures [49]. Because of their layered structure, 2D ferroelectric materials are ideal building blocks for FTJs. A recent study using $CuInP_2S_6$ as the tunnelling barrier and chromium/graphene as the asymmetric electrodes reports a TER of above $10^7$ [50]. This giant TER is attributed to a Fermi level shift as large as 1 eV by the polarization switching as shown in Figure 3a-b. Furthermore, 2D composite multiferroic tunnel junctions (MFTJs) incorporating both ferroelectric and ferromagnetic layers, $Fe_mGeTe_2/In_2Se_3/Fe_nGeTe_2$ (m, n = 3, 4, 5; m ≠ n) [51], have been studied theoretically. The multiple nonvolatile resistance states associated with polarization and magnetization alignment make them promising candidates for future non-volatile memory technologies.

**2D ferroelectric transistors and negative capacitance effect.** A ferroelectric field-effect transistor (Fe-FET) makes use of the spontaneous polarization of a ferroelectric gate layer to modulate the conductance of a semiconductive channel (such as doped silicon) [52]. However, the effect is hampered by imperfect interface due to structural and compositional incompatibility between conventional ferroelectric and semiconducting materials [53]. Recently, a 2D Fe-FET has been demonstrated using $CuInP_2S_6$ and $MoS_2$ as the gate dielectric and channel, respectively [54]. The device shows a large on/off ratio of ~$10^7$ and good retention performance, likely due to the high-quality interface between the 2D layers. Moreover, the coexistence of small band gap, high carrier mobility and switchable polarization in 2D ferroelectric semiconductors such as $In_2Se_3$ offers possibilities for novel ferroelectric transistors that combine the gate and channel layers in one. Si *et al* reported a large on/off ratio of above $10^8$, a maximum on current of 862 µA µm$^{-1}$ and a low supply voltage in a 2D ferroelectric transistor that uses α-$In_2Se_3$ as the channel as shown in Figures 3c-d [55].

Another opportunity offered by ferroelectric materials is the negative capacitance (NC) effect during the polarization reversal, which can effectively amplify the electric field applied to the channel and so decrease the subthreshold swing of an FET [56]. Recently, an average subthreshold swing (SS) lower than the Boltzmann limit has been demonstrated in a 2D CuInP$_2$S$_6$/ MoS$_2$ based FET [57], showing great potential for ultra-low-power applications. In both the tunnel junctions and Fe-FETs, the minimal surface reconstructi on and dangling bonds in 2D ferroelectrics presumably contributed to the superior device performance.

**2D ferroelectric-based spintronic devices.** The basic spin-orbit coupling (SOC) effect describes the interaction between spin and orbit angular momentum of an electron [58]. It has been shown that strong Rashba-type SOC exists in 2D ferroelectrics such as BiTeI and GeTe [59, 60], and coefficients one to two orders of magnitude larger than that in heavy metals and semiconductors have been predicted [59]. More importantly, the spontaneous polarization of ferroelectric materials is nonvolatile and switchable, ideal for electric-field control of spintronic devices. Recently, Varotto *et al* has demonstrated spin current injection from a magnetic Fe layer into a 2D ferroelectric GeTe layer [21]. The injected spin current is converted to a charge current, and more importantly, the sign of the charge current can be reversed by switching the ferroelectric polarization as illustrated in Figures 3e-f. 2D ferroelectric semiconductors/metals with strong SOC are therefore promising for next generation spintronic devices such as spin transistors and magnetic memories.

## Opportunities and challenges in 2D ferroelectrics

Following the above discussion of recent research in 2D ferroelectrics, we now offer some comments on the opportunities and challenges in the field.

**From 2D ferroelectrics to 2D multiferroics.** Multiferroics that possess both ferroelectricity and (anti)ferromagnetism have attracted much attention since magnetoelectric coupling enables electric-field control of magnetism, which may find applications in magnetic memory and spin transistors [35]. 3D single-phase multiferroics are mainly classified into two groups. In type-I multiferroics, e.g. BiFeO$_3$, the magnetic and ferroelectric orders are associated with the Fe and Bi ions, respectively [61]. Here Tc is relatively high for both the ferroelectric and the magnetic order separately, but the coupling between them is weak. In type-II multiferroics such as o-TbMnO$_3$, the ferroelectric order arises from spin interactions and displays strong magnetoelectric coupling, but the spontaneous polarization is very small with $T_C$ well below room temperature [36]. Inspired by earlier successes, recent theoretical studies have predicted a series of type-I ($CuCrX_2$ (X=Se or S) [62], bilayer $VS_2$ [63], $VOI_2$ [64], $(CrBr_3)Li$ [65]) and type-II ($Hf_2VC_2F_2$ [37]) 2D multiferroics. Experimental studies also revealed type-II multiferroicity in $NiI_2$ [25,26] and $CuCrP_2S_6$ [24], in which magnetic order induces electrical polarization. Also, combining charge redistribution induced ferroelectricity and magnetic ions in 2D systems could be another effective route to designing 2D multiferroics, circumventing the well known mutual exclusion between ferroelectricity and magnetism in traditional 3D systems. Overall, the reduced lattice dimensionality in 2D systems not only helps to stabilize long range magnetic order, but is also conducive to the survival of electric polarization in narrow band-gap and even metallic systems, offering tantalizing opportunities for 2D multiferroics.

The scarcity of single-phase 3D multiferroic materials has also driven researchers to look into composites of ferroelectrics and magnets [61]. In this respect, 2D vdW materials offer great potential. Their intrinsic layered structure means that different components can be stacked together like LEGO blocks without significant structural and chemical reconstruction at the interface, preserving their intrinsic properties down to atomic scale [5]. Even though mechanical coupling is relatively weak due to vdW interlayer interaction, charge redistribution and/or field effect can still occur, leading to 2D multiferroic heterostructures. Such 2D multiferroic composites have been predicted [51]. In principle, since they lack the lattice matching constraints of 3D heterostructures, 2D composite multiferroics have more degrees of freedom that can be exploited. For instance, one can stack them with a twist angle to create moiré multiferroics.

**Topological spin/polar structures in 2D systems and spin-based devices.** In magnetic materials that lack an inversion center, the SOC will produce a Dzyaloshinskii–Moriya interaction that tends to create a swirling spin configuration, resulting in arrays of magnetic skyrmions with identical chirality [66]. The skyrmions are topologically protected and can be moved by spin-polarized current, giving them potential for information storage and computing. However, currently the stabilization of skyrmion lattices usually occurs below room temperature and requires large magnetic fields [66]. Studies show that increased Rashba SOC coefficients in thin films greatly enhance the stability of magnetic skyrmions [67]. In this regard, 2D vdW multiferroics naturally possess two advantages, the intrinsic low lattice dimensionality and ferroelectric polarizations that can dramatically enhance their Rashba-type SOC coefficients. A sufficiently strong Dzyaloshinskii–Moriya interaction could therefore help to stabilize topological spin structures up to room temperature in 2D multiferroics.

Meanwhile, analogues of topological spin textures, that is, ferroelectric vortices and polar skyrmions have been proposed and observed in conventional 3D ferroelectric superlattices [68]. However, it remains an open question as to whether such chiral dipole structures can occur in 2D ferroelectric systems which have very different electromechanical properties. Interestingly, recent theoretical work predicts the possibility of noncollinear proper ferrielectricity and the electric analog of the Dzyaloshinskii–Moriya interaction that could promote chiral polarization textures in ferroelectrics [69,70]. The study of topological spin and /or polar textures is an active field with rich fundamental physics and technological potential, and extending into 2D ferroelectrics/multiferroics could bring more exciting discoveries.

Concerning of prototype devices, non-volatile control of spin-to-charge current conversion has been demonstrated in the 2D ferroelectric GeTe recently [21]. However, the ferroelectric control of spin precession and diffusion dynamics still needs further investigation. In other 2D ferroelectrics with theoretically predicted strong Rashba-type SOC, e.g. SnTe and BiTeI [59], the fundamental parameters such as spin Hall angle and spin diffusion lifetime and their correlation with ferroelectric polarization have not been discussed. The overall efficiency and reliability of spin manipulation using SOC effects as compared with other methods also require more theoretical and experimental investigation. Prototype devices such as spin transistors, memories and other spintronic devices based on 2D ferroelectrics/multiferroics remain to be realized.

**Polarization switching kinetics and reliability.** Fast and reliable polarization switching is crucial for ferroelectric-based devices. In conventional ferroelectrics, the switching process consists of domain nucleation, forward growth, and sideways growth stages, and can be categorized as either Kolmogorov-Avrami-Ishibashi (KAI)-type or nucleation-limited [71]. In contrast, polarization switching in 2D vdW ferroelectric materials can be convoluted with ionic conduction in ferroionic systems [72], sliding/twisting motions in stacking-driven ferroelectrics [19, 20] or carrier redistribution in narrow bandgap polar semiconductors/metals [13, 23], potentially leading to different switching kinetics. Furthermore, the discontinuity at the vdW gaps in 2D ferroelectrics may also affect the steps [29]. However, due to small volume and often large leakage in ultrathin samples, traditional polarization hysteresis measurement to study the kinetics is often challenging. Time and spatially resolved imaging and spectroscopies based on scanning probe microscopy should avoid these problems and allow studies of fatigue and retention behaviors in 2D ferroelectrics/multiferroics which are critical for applications.

An interesting example is the 2D ferroelectric α-$In_2Se_3$, which has a quintuple-layer structure. Here the simultaneous lateral and vertical displacements of middle layer Se atoms [10, 73, 74], with a structural change involving reorganization of covalent bonds, produces simultaneous switching of in-plane (nonlinear optical) and out-of-plane polarizations. This structural locking is believed to provide strong resistance to the depolarization field and be responsible for the high $T_C$ of ~700 K of this material [74]. Similarly, the polarization switching process in sliding and moiré ferroelectric systems, associated with a shear motion changing the interlayer stacking alignment, is expected to show very different kinetics [19]. Future improved understanding of the domain structure and switching mechanisms in these 2D ferroelectric systems could allow them to be employed for nonvolatile control of the electronic properties of 2D system.

**Bulk photovoltaic effect in 2D ferroelectrics.** The ferroelectric photovoltaic effect has been extensively studied because of the unique possibilities it could offer, including above band-gap open-circuit photovoltage, switchable photoresponse [75] and bypass the Shockley-Queisser limit [76] (although the power conversion efficiencies of ferroelectric solar cells remain very low). Nevertheless, 2D ferroelectrics are an excellent platform for exploring new physical effects in photo-electric conversion, such as related to symmetry breaking and phase transitions [77, 78].

Concerning applications, light polarization and energy sensitive photodetectors can be envisaged. 2D ferroelectrics are usually chalcogenides or halide compounds with more covalency in their chemical bonds, which result in larger band dispersions (small effective masses and high carrier mobilities) and narrower band gaps. Many 2D ferroelectrics, e.g. GeSe, SnTe [79], have band gaps close to that of silicon. These characteristics can improve light absorption and reduce photocarrier recombination providing a path to higher photoelectric conversion efficiency. 2D ferroelectrics may thus have advantages over 3D counterparts in optoelectronic devices [77, 80].

**Large-scale 2D ferroelectric thin film preparation.** Most of the current studies on 2D ferroelectrics focus on fundamental physics and laboratory-scale device demonstrations with micrometer-sized samples but wafer-scale fabrication of single-crystalline thin films by either *in-situ* growth or *ex-situ* transfer, will be necessary for real-life applications. Chemical vapor deposition (CVD) remains the most highly developed and successful approach for mass production of 2D materials, in terms of uniformity and crystallinity of the films, cost effectiveness, and scalability [81]. In the epitaxial growth of conventional oxide ferroelectric thin films, the growth kinetics are governed by the strong ionic and/or covalent bonding between the films and atomically flat substrates which enforce the structure of the film. In stark contrast, the weak vdW interactions between 2D ferroelectrics and the substrates usually result in nucleation-controlled polycrystalline films [82]. To facilitate oriented or less disordered growth, anisotropic or vicinal substrates with atomic steps can be exploited as natural templates, echoing the strategies used in domain engineering of oxide ferroelectrics [83]. For example, by using metallic Cu foils as substrates, wafer-scale single-crystal graphene and h-BN have been achieved [84, 85]. Using a seed-assisted growth process, single-crystal and uniform 2H-MoTe$_2$ thin films have been fabricated on 1-inch amorphous and insulating wafer recently [86].

However, the range of 2D materials so far grown by CVD in large-scale films is currently limited, with none yet being ferroelectric. Furthermore, in CVD it is challenging to obtain the precise control of vapor concentration needed to grow multi-element complex compounds growth using CVD. Other deposition techniques can also be explored, such as molecular beam epitaxy (MBE), metal-organic chemical vapor deposition (MOCVD), atomic layer deposition (ALD) and pulsed laser deposition (PLD). The recent reports of large-scale growth of black phosphorus film by PLD [87] and ferroelectric α-$In_2Se_3$ by MBE [88] are encouraging in this regard.

To summarize, we list the main challenges in 2D ferroelectric research in Box 1, and hopefully addressing these issues would prove fruitful in the coming years.

## Outlook

The past few years have witnessed an explosion of research activities in 2D ferroelectrics. New mechanisms for spontaneous polarization in 2D systems have been stablished. The reduced lattice dimensionality enables ferroelectric semiconductors and even ferroelectric metals. The vdW gaps not only change the electromechanical behaviors of 2D ferroelectrics, but also enable high quality heterostructures free of interfacial defects and epitaxy of large-area thin films.

Looking forward, as new aspects of ferroelectric materials and physics continue to be uncovered and insights gained, we expect to see device concepts making use of the unique or enhanced properties of 2D ferroelectrics/multiferroics being developed and tested. The flourishing of 2D ferroelectrics will expand the catalog of long-range order and coupling phenomena in low-dimensional systems and creates new possibilities for next-generation nanoelectronics and spintronics.

**Figures.**

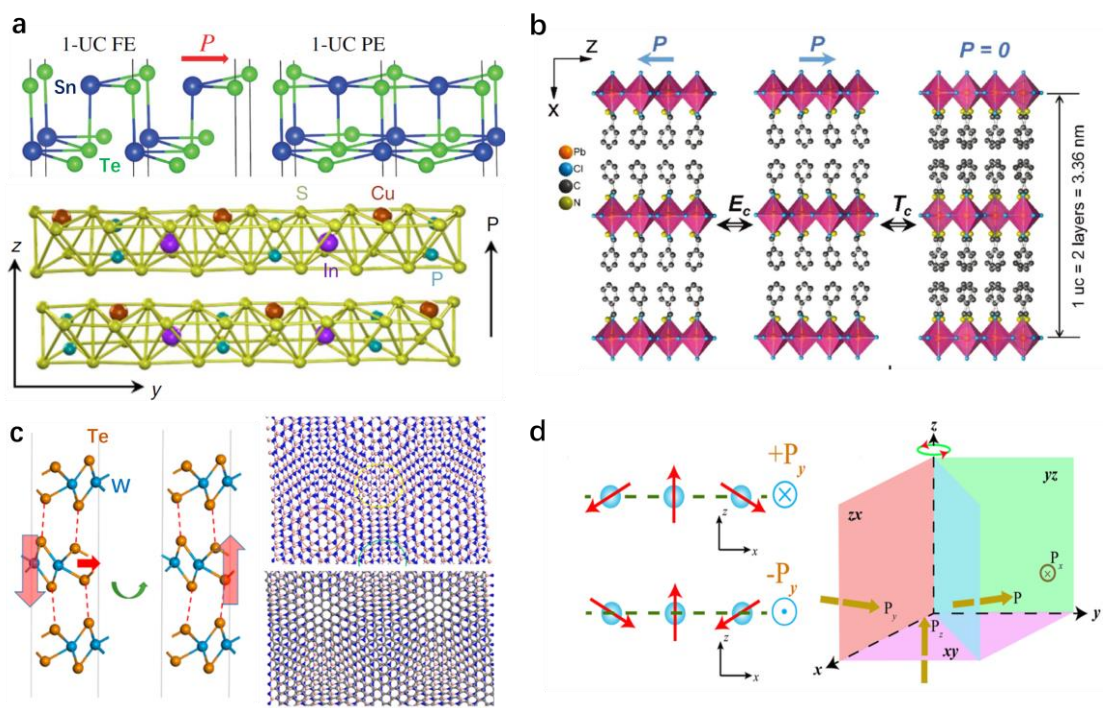

**Fig. 1: Different mechanisms of spontaneous polarization in 2D ferroelectrics. a,** Noncentrosymmetric crystal structure with polarization induced by Sn and Cu displacements in SnTe (top) and CuInP$_2$S$_6$ (bottom), respectively. **b,** Polar molecular group induced polarization in BA$_2$PbCl$_4$ below T$_C$ (left and middle) and the order-disorder transition across T$_C$ (right). **c,** Charge redistribution and interlayer sliding induced polarization in trilayer WTe$_2$ (left, the large pink arrows indicate the polarizations), and moiré ferroelectricity in small angle twisted BN bilayer (right top) and graphene/BN heterostructure (right bottom). **d,** Spin texture induced polarization in Y-type antiferromagnetic Hf$_2$VC$_2$F$_2$. The helical clockwise (left top) and counterclockwise (left bottom) ordered spins generate polarizations perpendicular to the spin planes. Rotation of the spin plane and the corresponding polarization along the z axis is shown on the right. Figures reproduced with permission from: **a,** ref. [8,28], Springer Nature Limited, American Physical Society; **b,** ref. [12], WILEY-VCH Verlag GmbH & Co.; **c,** ref. [14, 34], American Chemical Society; **d,** ref.[37], American Chemical Society.

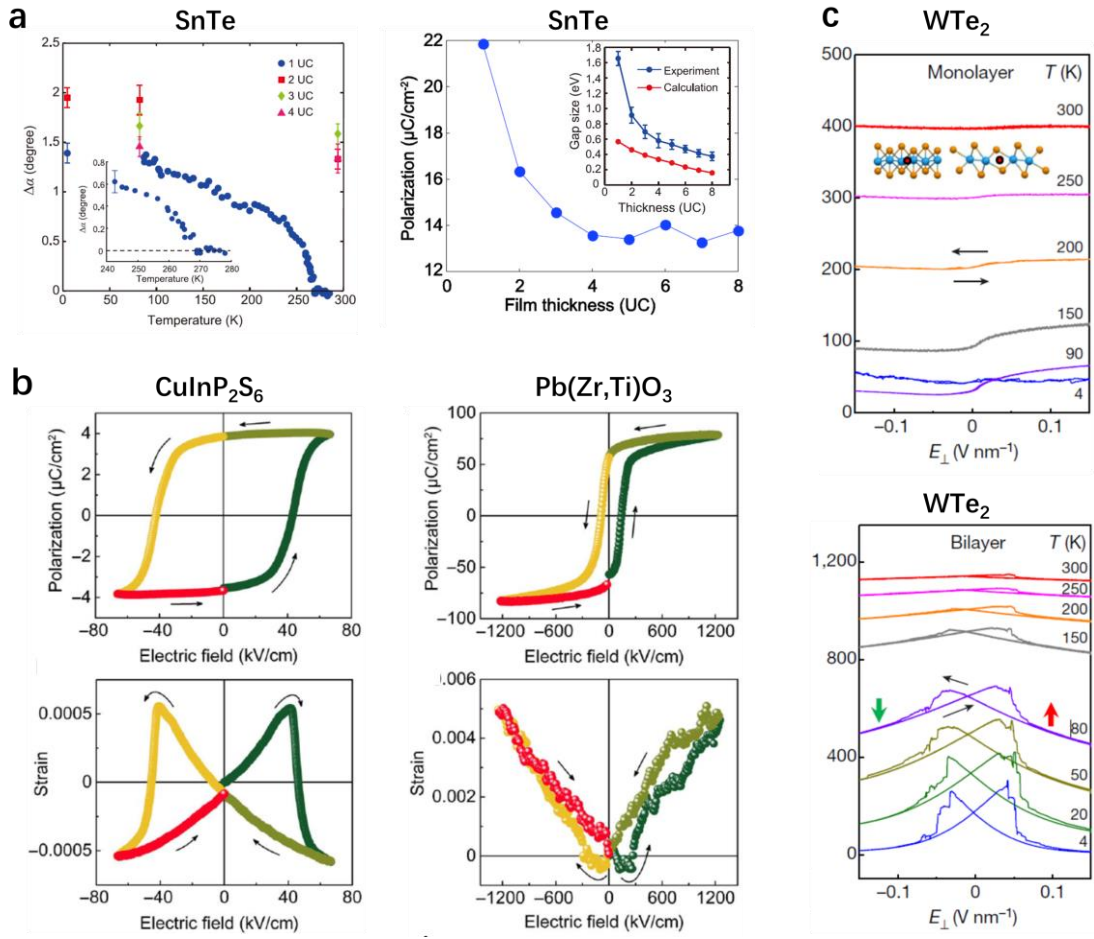

**Fig. 2: Effects of reduced lattice dimensionality on ferroelectricity in 2D systems. a,** Temperature dependence of the distortion angle (proportional to the square of polarization) in SnTe thin flakes (left). The inset shows the second-order phase transition of 1 unit-cell (UC) SnTe around 270 K; The calculated spontaneous polarization increases with decreasing SnTe thickness (right). The inset exhibits the evolution of band gap with sample thickness. **b,** The negative (left) and positive (right) longitudinal piezoelectric response of CuInP$_2$S$_6$ and Pb(Zr,Ti)O$_3$, respectively. **c,** The conductance of flakes of monolayer (top)- and bilayer (bottom)-WTe$_2$ as perpendicular electric field $E_\perp$, applied between upper and lower graphite gates separated by thin h-BN sheets, is swept (black arrows), respectively; the red and green arrows indicate the polarization direction, corresponding to the two states drawn in Fig. 1**c**. Figures reproduced with permission from: **a,** ref. [9], American Association for the Advancement

of Science; **b,** ref. [41], American Association for the Advancement of Science; **c,** ref. [13], Springer Nature Limited.

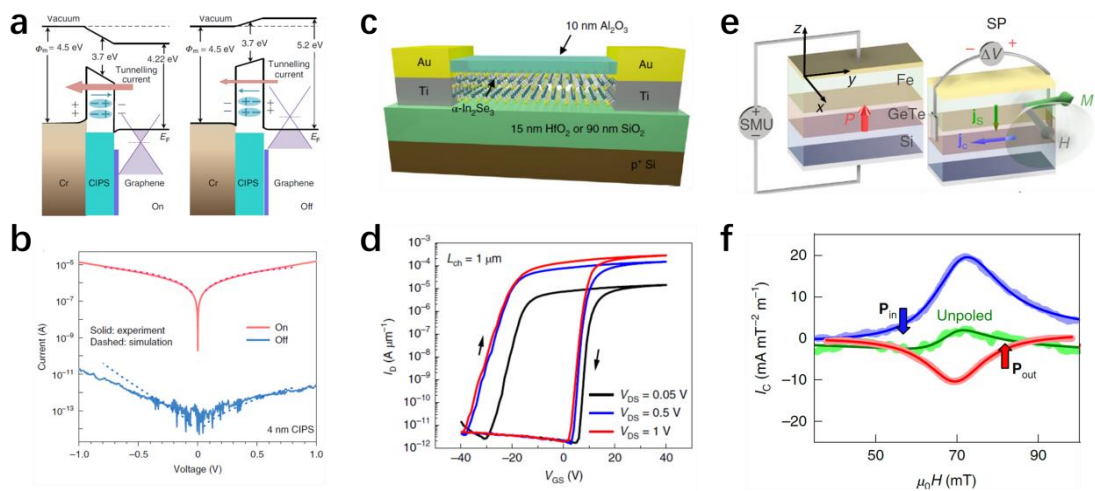

**Fig. 3: Prototype devices based on 2D ferroelectrics. a,** The tunnelling barrier is tuned by polarization switching (cyan arrows) in a CuInP$_2$S$_6$ FTJ, leading to **b,** a TER of above 10$^7$. **c,** Schematic of a novel 2D Fe-FET using ferroelectric α-In$_2$Se$_3$ as the channel layer, and **d,** $I_D$–$V_{DS}$ characteristics controlled by the polarization switching (black arrows) of α-In$_2$Se$_3$ at room temperature. **e,** Schematic of the experimental setup of ferroelectric switching controlled spin to charge current conversion in GeTe. **f,** Changes in converted charge currents controlled by the polarization switching of GeTe. Figures reproduced with permission from: **a-b,** ref. [50], Springer Nature Limited; **c-d,** ref. [55], Springer Nature Limited; **e-f,** ref. [21], Springer Nature Limited.

**Box 1: Challenges in 2D ferroelectric research.**

    **2D multiferroics:** achieving robust 2D type-I, type-II and composite multiferroics with high $T_C$s. Exploring the underlying mechanisms of magnetoelectric coupling in 2D systems (panel **a**). **Topological spin/polar structures in 2D systems:** enabling topological polar and/or spin textures in 2D ferroelectrics/multiferroics and understanding the key factors that stabilize topological structures in low-dimension ferroic systems (panel **b**). **Spin-based devices in 2D systems:** Designing and fabricating prototype spin devices with excellent performance, e.g., low-energy consumption, high density, and long durability based on 2D ferroelectrics/multiferroics (panel **c**). **Polarization switching kinetics and reliabilities:** reveal of ultrafast kinetics of polarization switching in 2D ferroelectrics and investigating the fatigue and retention behaviors in 2D ferroelectrics/multiferroics (panel **d**). **Bulk photovoltaic effect in 2D ferroelectrics:** unravelling new physical phenomena in high-efficiency photo-electric conversion and developing new photoelectric devices based on 2D ferroelectrics (panel **e**). **Large-scale 2D ferroelectric thin film preparation:** developing new wafer-scale growth methods such as substrates modification, seed-assisted crystalline process, and other deposition techniques etc. suitable for 2D ferroelectric thin films with uniformity (panel **f**).

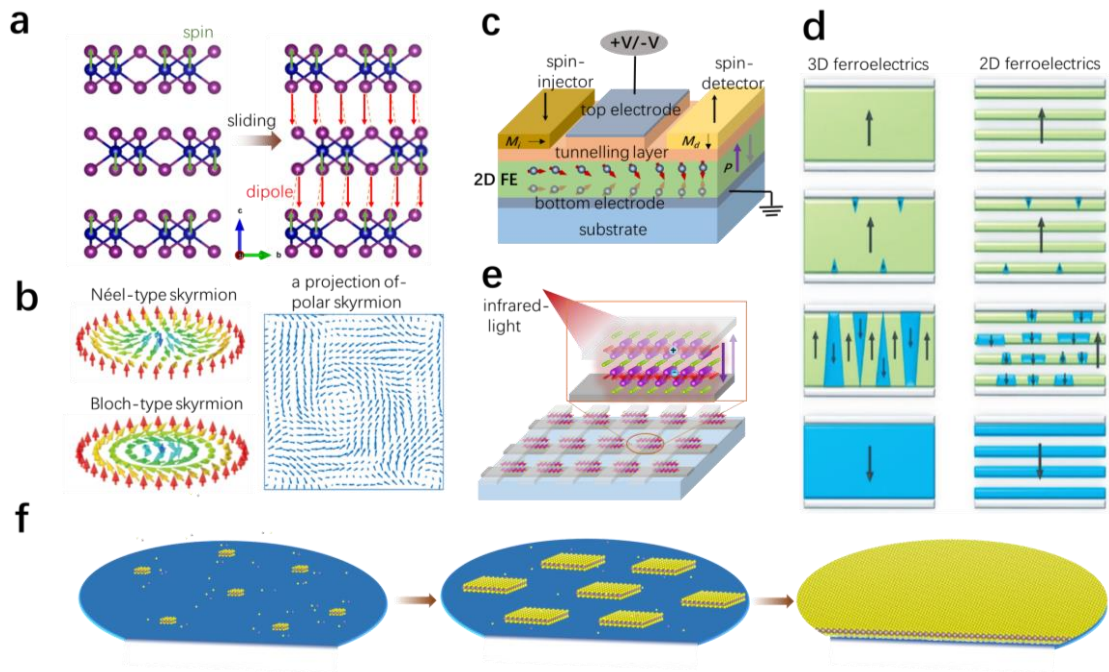

Panels reproduced with permission from: **b,** ref.[66, 68], Springer Nature Limited; **c,** ref.[60], Springer Nature Limited; **d,** ref.[29], WILEY-VCH Verlag GmbH & Co.

**Acknowledgements**

L.Y. acknowledges financial support from National Natural Science Foundation of China under Grant No. 12074278, the Natural Science Foundation of the Jiangsu Higher Education Institution of China under Grant No. 20KJA140001, Priority Academic Program Development (PAPD) of Jiangsu Higher Education Institutions and Jiangsu Specially-Appointed Professors Program. L.Y. also acknowledges the support from the National Natural Science Foundation of China (11774249, 12074278), the Natural Science Foundation of Jiangsu Province (BK20171209), and the Key University Science Research Project of Jiangsu Province (18KJA140004, 20KJA140001). J.W. acknowledges supports from the National Natural Science Foundation of China (Grant No. 12074164), Guangdong Provincial Key Laboratory Program (2021B1212040001) from the Department of Science and Technology of Guangdong Province, and the startup grant from the Southern University of Science and Technology (SUSTech), China